\documentclass[twocolumn, letter]{emulateapj}
\usepackage{multirow}
\usepackage{graphics}
\usepackage{amsmath, amsthm, amssymb}
\usepackage{epsfig}
\usepackage[hidelinks]{hyperref}
\usepackage{natbib}

\usepackage{graphicx}
\usepackage[dvipsnames]{xcolor}

\begin{document}

\title{
On the stellar velocity distribution in the solar neighborhood in the light of Gaia DR2
}


\author{Tatiana A. Michtchenko$^{1}$}\email[e-mail: ]{tatiana.michtchenko@iag.usp.br}

\author{Jacques R.\,D. L\'epine$^{1}$}\email[e-mail: ]{jacques.lepine@iag.usp.br}

\author{Angeles P\'erez-Villegas$^{1}$}\email[e-mail: ]{mperez@iag.usp.br}

\author{Ronaldo S. S. Vieira$^{1}$}\email[e-mail: ]{rss.vieira@usp.br}

\author{Douglas A. Barros$^{2}$}\email[e-mail: ]{douglas.barros@alumni.usp.br}

\affiliation{$^1$Universidade de S\~ao Paulo, IAG, Rua do Mat\~ao, 1226, Cidade Universit\'aria, 05508-090 S\~ao Paulo, Brazil\\
$^2$Rua Sessenta e Tr\^es, 125, Olinda, 53090-393 Pernambuco, Brazil}

\date{\today}

\begin{abstract}
The aim of this work is to contribute to the understanding of the stellar velocity distribution in the solar neighborhood (SN). We propose that the structures on the $U$--$V$ planes, known as the moving groups, can be mainly explained by the spiral arms perturbations. The applied model of the Galactic disk and spiral arms, with the parameters defined by observational data and with pattern speed $\Omega_p=$28.0\,km\,s$^{-1}$\,kpc$^{-1}$, is the same that allowed us to explain the origin of the Local Arm and the Sun's orbit trapped inside the corotation resonance (CR). We show that the $U$--$V$ picture of the SN consists of the main component, associated with the CR, and the inner and outer structures, which we could  associate with the Hercules and Sirius streams, respectively. The Coma-Berenices and Hyades-Pleiades groups and the Sun itself belong to the main part.  The substructures of Hercules are formed mainly by the nearby 8/1, 12/1, and even 6/1 inner Lindblad resonances, while Sirius is shaped by the bulk of overlapping outer Lindblad resonances, -8/1, -12/1, -16/1, which are stuck to the CR. This richness in resonances only exists near corotation, which should be of the spiral arms, not of the Galactic bar, whose stable corotation zone is far away from the Sun. The model's predictions of the velocity distribution match qualitatively and quantitatively the distribution provided by Gaia DR2.
\end{abstract}


\maketitle

\section{Introduction}
\label{intro}

In recent years, great efforts have been dedicated to give plausible explanations to the moving groups and the rough bimodality which are observed in the velocity distribution of the solar neighborhood (SN) \citep[][]{dehnen2000AJ, quillenMinchev2005AJ, antojaEtal2014AA}.
However, these kinematic structures are not yet settled and it seems that, at present, we lack models which could explain the observations as complete as those presented by Gaia DR2 \citep{Gaia2018A}.

The stellar velocity distribution near the Sun shows density clumps, which were first related to the disruption of open clusters (see \citealp{antojaEtal2010LNEA}, for a review). Nevertheless, the heterogeneity of the ages of these groups  \citep{bensbyEtal2007ApJL,famaeyEtal2008AA} was in contradiction with this hypothesis, demanding a new scenario. Furthermore, a question, which has drawn attention, was the apparent bimodality in the $U$--$V$ distribution, in which one of these groups, Hercules, is separated from the main component of the velocity distribution in the SN.

Some dynamical scenarios explain this separation considering it as an imprint of the Galactic bar perturbations \citep[][]{dehnen1999ApJL, dehnen2000AJ, bovy2010ApJ, antojaEtal2014AA, perezvillegasEtal2017ApJL}. Others study spiral arms perturbations \citep{desimoneEtal2004MNRAS,quillenMinchev2005AJ,antojaEtal2011MNRAS,quillenEtal2018arXiv} or a mix of bar+spirals \citep{antojaEtal2009ApJL}.
Additionally, transient spiral structures and phase wrapping were recently proposed to explain $U$--$V$ distribution  in the SN \citep{Hunt2018arXiv180602832H,antojaEtal2018arXiv}.

Our explanation presented here is based on strong observational evidence, which include the Sagittarius-Carina and Perseus spiral arms,  the Local Arm \citep[][hereafter Paper-I]{LepineEtal2017ApJ} and the proximity between the spiral corotation and the solar circle \citep{mishurovZenina1999AA}. We find that the complex structure of the $U$--$V$ plane in the SN can be represented roughly by three components: the main component, associated with the spiral corotation resonance (CR), Hercules, associated with inner Lindblad resonances (ILRs), and Sirius, associated with the bulk of overlapping outer Lindblad resonances (OLRs). Both ILRs and OLRs are high-order resonances, which naturally appear in the corotation neighborhood.


\begin{figure*}[h]
\begin{center}
\epsfig{figure=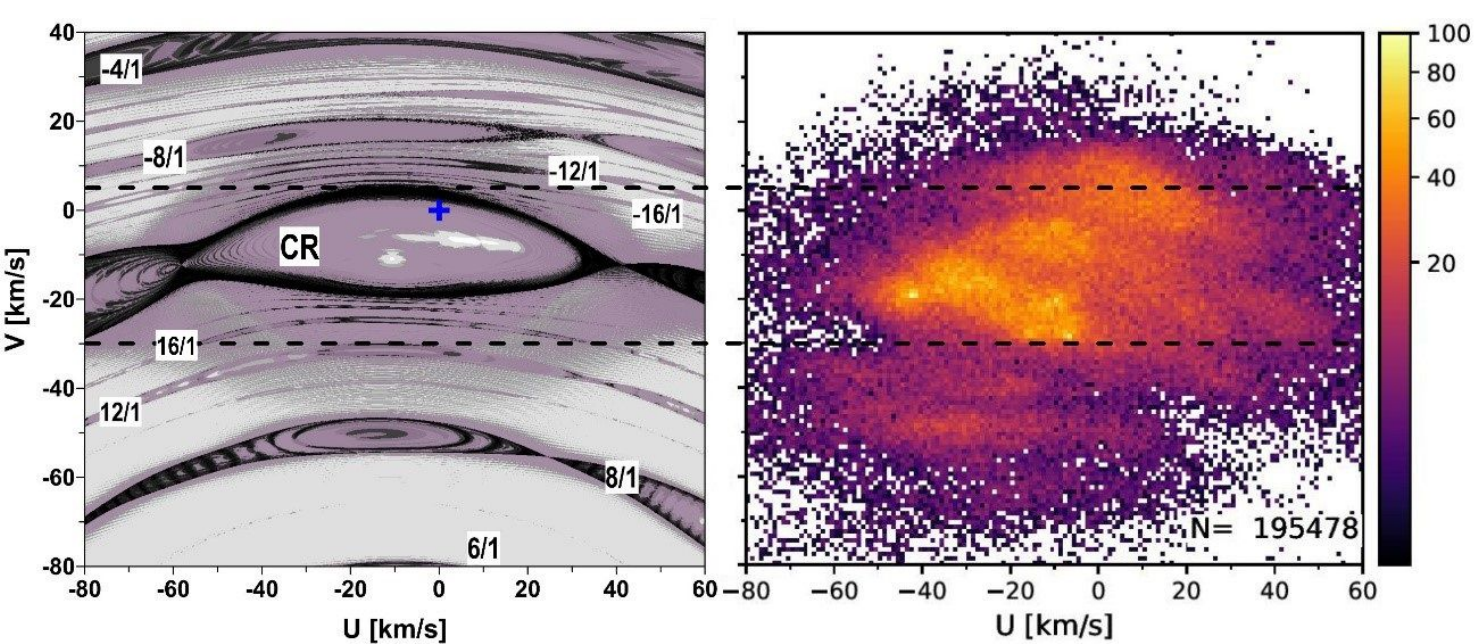,width=1.0\textwidth,angle=0}
\caption{Left: Dynamical map on the $U$--$V$ plane constructed at the initial position of the Sun, $R=8.0$\,kpc and $\varphi=90^\circ$ (blue cross).  The main resonances are identified by the corresponding ratios. The dashed lines separate roughly the main component from the Hercules (V$<$-30\,km/s) and Sirius (V$>$5\,km/s) regions.
Right: Heliocentric velocity distribution of the Gaia stars within 150\,pc from the Sun as a 2D histogram of the velocity with a bin-size of 1\,km\,s$^{-1}$ in both directions. The color scale indicates the percentage of stars per (km/s)$^2$.
}
\label{fig:fig1}
\end{center}
\end{figure*}

\section{Data \& Model}
\label{data}

{\it Gaia} mission recently released its second set of data \citep{Gaia2018A}, which provides 6D phase-space coordinates for 7,224,631 stars, i.e. positions in the sky $(\alpha,\delta)$, parallaxes $\varpi$, proper motions $(\mu_{\alpha}^*,\mu_{\delta})$ \citep{lindergrenEtal2018AA}, and radial line-of-sight velocities \citep{KatzEtal2018A} for stars with $G<$13\,mag. We restrict ourselves to stars with parallax errors smaller than $20\%$. Additionally, we focus on stars inside 1\,kpc from the Sun, where using $d=1/\varpi$ as distance estimate is acceptable. As result, our final catalog contains 3,105,498 stars. In order to convert the positions, parallaxes, proper motions on the sky, and radial velocities of the stars into Cartesian Galactic phase-space positions and velocities, we use the \texttt{galpy} python tools \citep{bovy2015ApJS}.

Our Galactic model is composed of an axisymmetric background derived from the observed rotation curve \citep[][]{BarrosLepine2016},  and a four-arm spiral structure, which has Gaussian-shaped azimuthal profiles \citep[][]{junqueiraEtal2013AA,Michtchenkoetal2017AA}.
The spiral arms profile, in the rotating frame, is given by
\begin{equation}
\label{eq:Phi_s}
 \Phi_{\rm sp}(R,\varphi) = - \zeta_0\,R\,e^{-\frac{R^2}{\sigma^2}[1-\cos(m\varphi-f_m(R))]-\varepsilon_s R},
\end{equation}
where $R$ is the Galactocentric distance and $\varphi$ is the azimuthal angle, measured with respect to the $X-$axis perpendicular to the Sun's direction. The parameters are: the spiral arm strength $\zeta_0=200.0$\,km$^2$\,s$^{-2}$, the width $\sigma\sin i=10.0\times\sin(15^\circ)$\,kpc ($i$ is the pitch angle), and the radial scale-length $\varepsilon_s^{-1}=4.0$\,kpc, while $f_m(R)$ is a four-arm logarithmic shape function.
More details about the parameters are presented in Paper-I. The arm's width was slightly increased, reducing the chaotic layers of the CR that allows us to clearly visualize the high-order resonances, close to the CR.

Our recent study  (Paper-I) shows that the origin of the Local Arm can be explained by the proximity of the Sun to the CR \citep[][]{diasLepine2005ApJ}, provided the $\Omega_p$--value is in the range 27$\pm$2\,km\,s$^{-1}$\,kpc$^{-1}$. We choose here $\Omega_p=28$\,km\,s$^{-1}$\,kpc$^{-1}$, corresponding to a corotation radius at 8.2\,kpc. In our model, the Sun is placed at $R_0=8.0$\,kpc, with a circular velocity of $V(R_0)=230$\,km\,s$^{-1}$, assuming a solar peculiar motion with respect to the circular orbit of $(U_0,V_0) = (11.1,12.24)$\,km\,s$^{-1}$ \citep{schonrichBinneyDehnen2010MNRAS}.

\section{Velocity distribution in the SN}
\label{SN}

Using our model, we calculate the dynamical map on the $U$--$V$ plane, presented in Fig.\,\ref{fig:fig1}\,left, for $R=8.0$\,kpc and $\varphi=90^\circ$. Lighter grey tones on the map show the regular orbits, while increasingly dark tones show the increasing instabilities and chaotic motion. The resonances are recognized as structures surrounded by chaotic layers, associated to resonance separatrices. The purple color highlights the resonant and quasi-resonant domains (or zones of resonance influence). For the calculations, we use integration times of $\sim$12\,Gyrs; the structures appear clearly on the maps after 2\,Gyrs ($\sim10$ orbits around the galactic center), but we perform longer integrations to get better defined separatrices.  The $U$--$V$-velocity distribution within 150\,pc from the Sun provided by Gaia DR2 is shown in right panel of Fig.\,\ref{fig:fig1}.

The initial comparison of the dynamical map (left) with the density distribution (right) in Fig.\,\ref{fig:fig1} can be done only qualitatively, since the positions of the resonances on the $U$--$V$ plane depend on $R$. We observe a dominant presence of the CR in the central part of the plane and identify the well-known moving groups, Coma-Berenices and Hyades-Pleiades, as objects belonging to the CR and its zone of influence. The abrupt cut-off on the upper left-side of the main region, observed previously in the data \citep{antojaEtal2008AA}, would be explained by the corotation separatrix. The CR separates the $U$--$V$ plane in the inner and outer regions. The inner region, populated by several high-order ILRs (8/1 being the most influent), is  associated to Hercules. Lindblad resonances are also present in the outer region, which we associate to Sirius.  The detailed   comparison of the $U$--$V$-structures and Lindblad resonances is presented in Sect.\,\ref{Hercules}.

\begin{figure}[h]
\begin{center}
\epsfig{figure=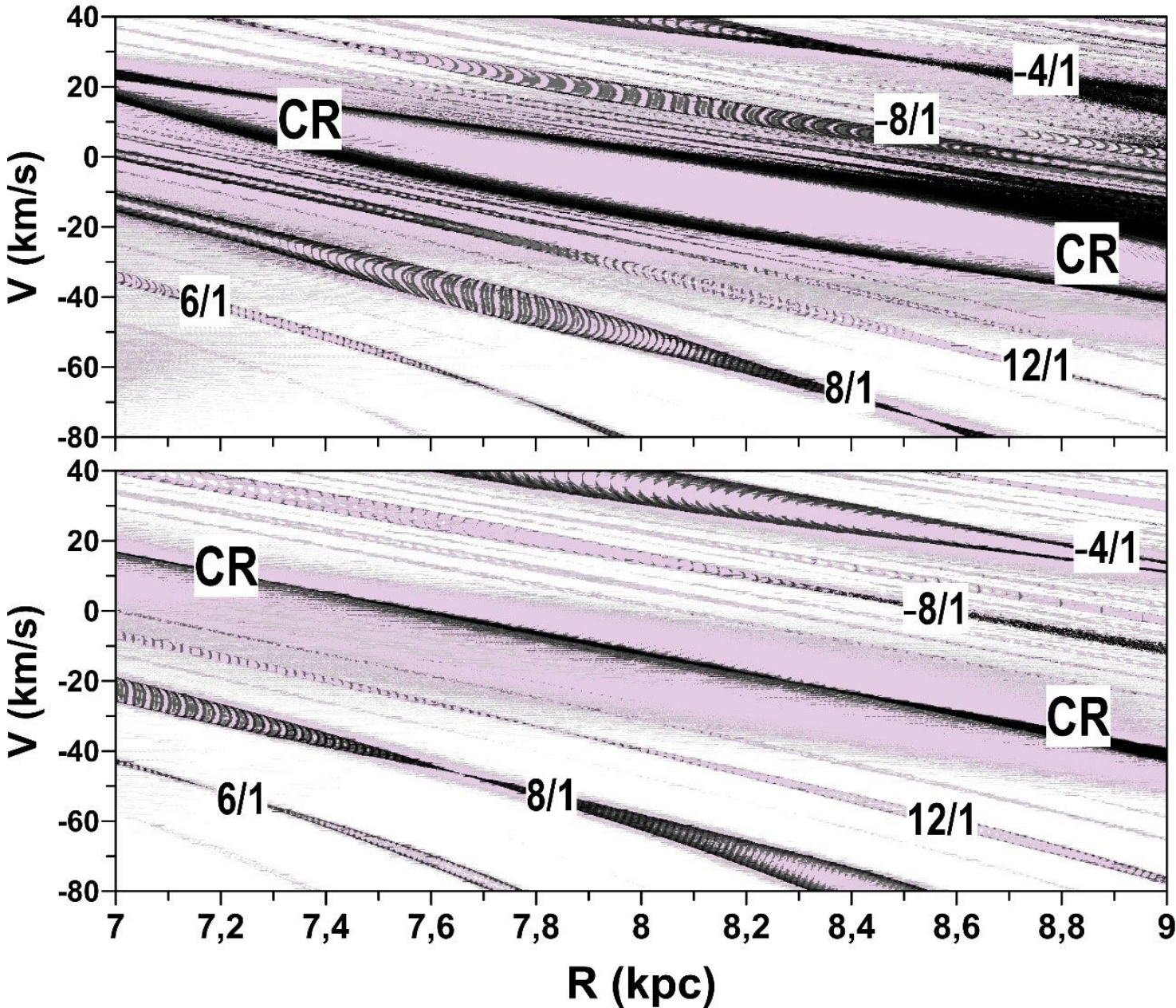,width=0.99\columnwidth,angle=0}
\caption{Dynamical maps of the CR and nearby resonances as functions of $R$, on the $R$--$V$ planes ($\varphi=90^\circ$), calculated for $U=-30$\,km\,s$^{-1}$ (top) and $U=+40$\,km\,s$^{-1}$ (bottom).
}
\label{fig:fig2}
\end{center}
\end{figure}

\section{Velocity distribution as a function of $R$}
\label{results}

The connection between the corotation/resonances and the moving groups is verified regarding  the main features of resonant motion. For instance, our model predicts that the resonant $V$-velocities will depend on Galactocentric distance, as shown in Fig.\,\ref{fig:fig2}, where we present dynamical maps on the $R$--$V$ planes, for two values of $U$, $-30$\,km\,s$^{-1}$ (top) and $+40$\,km\,s$^{-1}$ (bottom). On both planes, the resonant $V$-velocities continuously decrease with increasing $R$; consequently,  the resonance pattern in Fig.\,\ref{fig:fig1}\,left (calculated for $R=8.0$\,kpc) will move up/down for smaller/larger $R$--values. The apparent stable domains of resonant objects on the $U$--$V$ plane are broader than those shown in Fig.\,\ref{fig:fig1}\,left, since the radial distances of the sample are distributed inside 150\,pc from the Sun in Fig.\,\ref{fig:fig1}\,right.

Absence of objects is expected in the vicinity of the resonance saddle points, where the stable zone is of zero-width and chaotic motion dominates, as shown in Fig.\,\ref{fig:fig2}. The positions of the saddle points and, consequently, of the low-density regions on the $U$--$V$ planes change with radius. In addition, we can see that, for $U=-30$\,km\,s$^{-1}$ (top panel), the widths of the stable CR and 8/1\,ILR are non-zero for all considered $R$- and $V$-values, excluding only very low $V$-values.  In contrast, for $U=+40$\,km\,s$^{-1}$ (bottom panel), the width of the CR is nearly zero over the whole $R$--$V$ plane, while the width of the 8/1\,ILR drops to zero in the domains 7.4$<$R$<$8.5\,kpc and -80$<$V$<$-30\,km/s. The distances between the CR and nearby ILRs increase with increasing Galactocentric distances.
\begin{figure}[h]
\begin{center}
\epsfig{figure=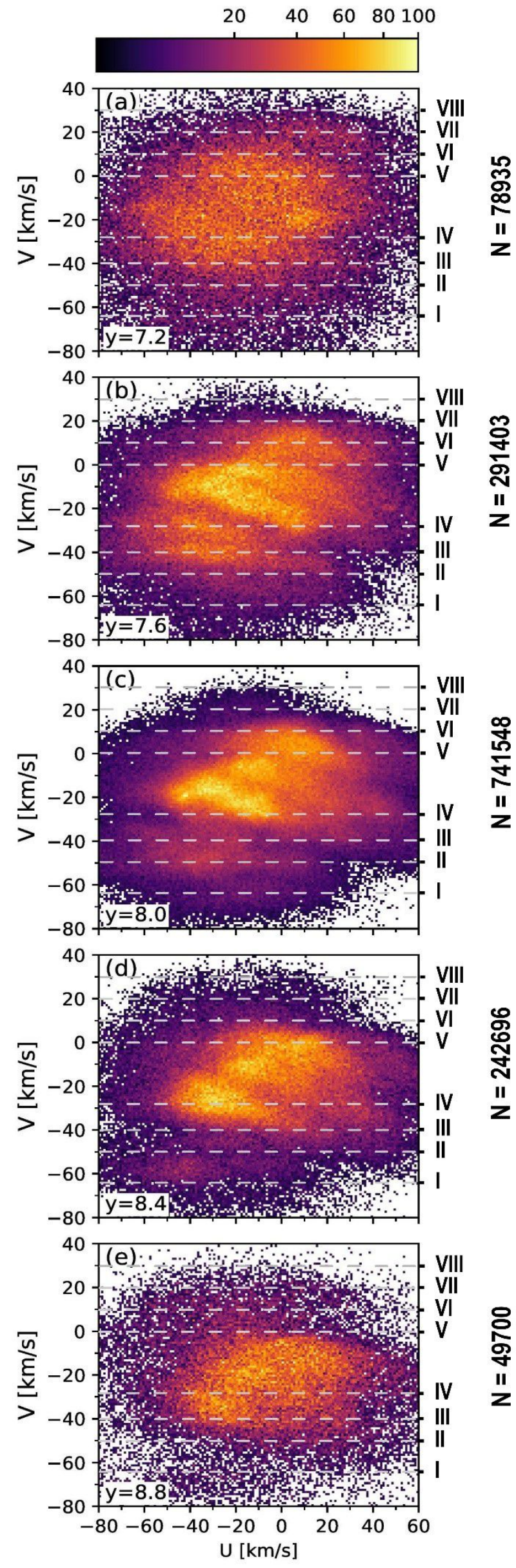,width=0.79\columnwidth ,angle=0}
\caption{Density portraits of the $U$--$V$--plane, as functions of Galactocentric distance; five consecutive zones are labeled from (a) to (e). Each panel shows the velocity distribution in a square of 400\,pc on each side. The center of each bin in Cartesian coordinates is indicated in the bottom left corner, with $X=0$ for all bins. The color scale is the same as of Fig.\,\ref{fig:fig1}\,right. Constant-$V$ dashed lines are denoted as I(-66), II(-50), III(-40), IV(-27), V(0),VI(+10), VII(+20), VIII(+30) km/s. }
\label{fig:fig3}
\end{center}
\end{figure}

The main features of the CR and Lindblad resonances predicted by our model are compared with the $U$--$V$ stellar velocity distribution in Fig.\,\ref{fig:fig3}, where we show five planes with square bins of 400\,pc. The center of each bin, in Cartesian coordinates with respect to the Galactic center, is $(X,Y)= (0,7.2)$, $(0,7.6)$, $(0,8.0)$, $(0,8.4)$, and $(0, 8.8)$, in kpc. All planes exhibit a dominant presence of the main component associated with effects of the CR, which slowly scrolls down on the $U$--$V$ plane with increasing R, as predicted in Fig.\,\ref{fig:fig2}.

\begin{figure*}[h]
\begin{center}
\epsfig{figure=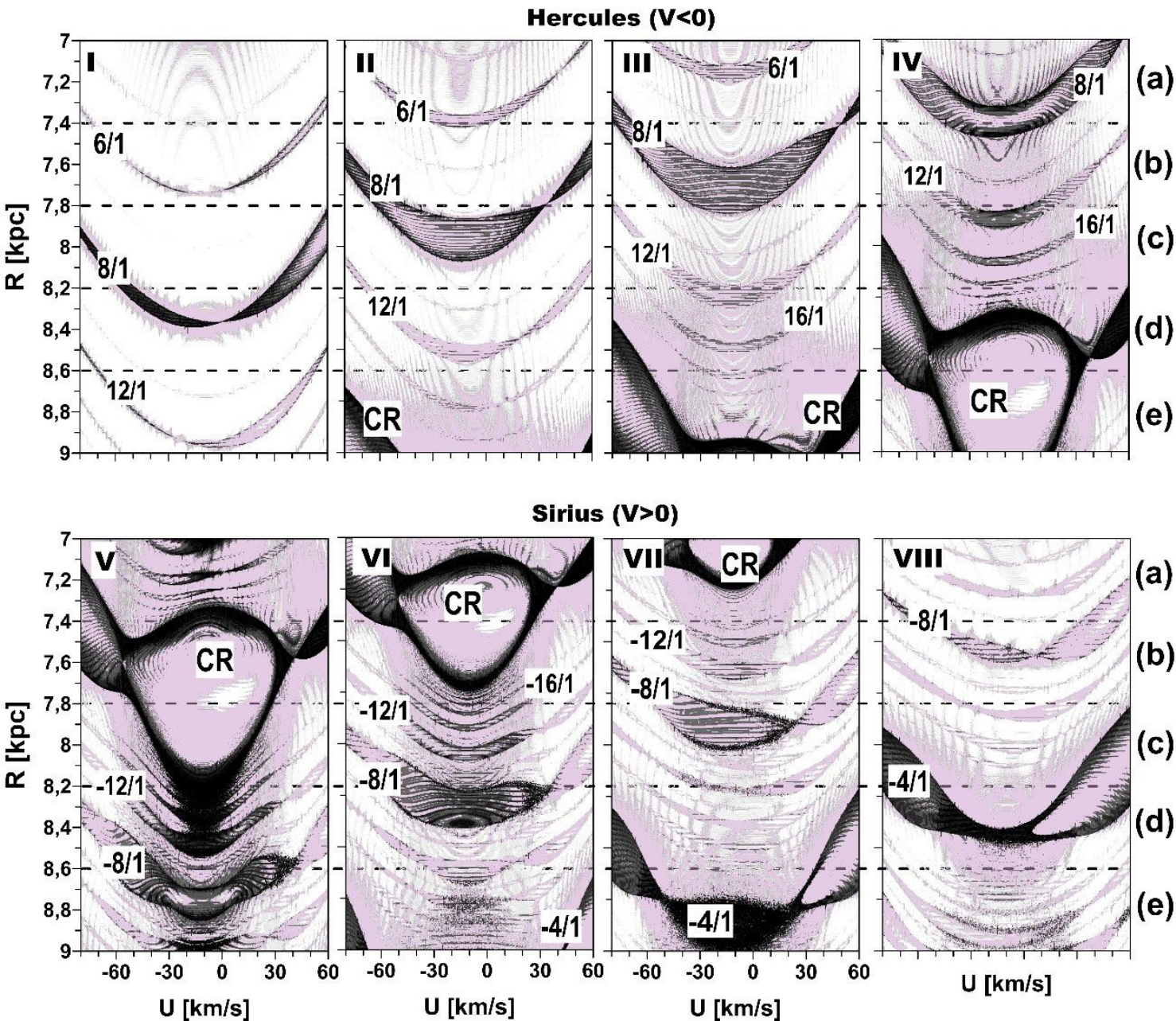,width=0.99\textwidth ,angle=0}
\caption{Dynamical maps of the $U$--$R$ planes ($\varphi=90^\circ$), for the $V$-levels corresponding to the horizontal lines I to VIII of Fig.\,\ref{fig:fig3}. Horizontal dashed lines cut the space in five bins, from (a) to (e), corresponding to those presented in Fig\,\ref{fig:fig3}. The ILRs and OLRs are labeled by the corresponding positive and negative ratios, respectively.}
\label{fig:fig4}
\end{center}
\end{figure*}

\section{Hercules' and Sirius' substructures}
\label{Hercules}

Gaia DR2 reveals, for the first time, a very complex structure of the inner and outer zones on the $U$--$V$ plane, associated to the Hercules and Sirius streams, respectively. In order to compare the results of our model with the observational data of Fig.\,\ref{fig:fig3}, we present in Fig.\,\ref{fig:fig4} the dynamical maps on the $U$--$R$ planes, for eight $V$-levels labeled from I to VIII, according to the $V$-values shown by horizontal lines in Fig.\,\ref{fig:fig3}. Each panel is further divided in five $R$-bins, which match the five planes in Fig.\,\ref{fig:fig3}, from (a) to (e), with the corresponding mean $\bar{R}$=y.

The main features we observe in Fig.\,\ref{fig:fig4}  are the corotation and Lindblad resonances,
which shift continuously towards smaller $\bar{R}$ for increasing $V$ (a behavior already observed in Fig.\,\ref{fig:fig2}). Consequently, the CR dominates the region between the IV- and V-levels on the maps in Fig.\,\ref{fig:fig4}. Each resonance has a specific topology of its domain around stable center and saddle points, which is asymmetric with respect to the $U$-component. Assuming that these dynamical features give rise to structures in the stellar velocity distribution, we look for their identification in Fig.\,\ref{fig:fig3}. We expect to observe the maximal density of the resonant population the resonance domain and the minimal density in the proximity of the saddle points.

\subsection{Hercules}

Since the Hercules stream is an inner structure, we select four negative $V$-levels below the CR (I to IV  in Fig.\,\ref{fig:fig3}) to characterize this group. All corresponding panels in Fig.\,\ref{fig:fig4}\,top show the presence of the ILRs, from which 8/1 is strongest. Far from the Sun, at $\bar{R}=7.2$\,kpc, 8/1\,ILR already appears at the IV-level. Accompanied by the weaker 6/1\,ILR, it  produces an $U$-$V$-substructure between the III- and IV-levels, visible in Fig.\,\ref{fig:fig3}(a). Approaching to the Sun, at $\bar{R}=7.6$\,kpc, this substructure becomes very dense and splits in two horizontal branches, stronger at III-level and weaker at IV-level, clearly observed in Fig.\,\ref{fig:fig3}(b). The corresponding radial bin, (b) in Fig.\,\ref{fig:fig4}, associates the stronger branch to the 8/1\,ILR, with its $U$-extension ranging approximately from -70\,km\,s$^{-1}$ to 30\,km\,s$^{-1}$, as limited by the proximity of the 8/1\,ILR saddle points. The weaker branch is associated, in panel IV(b) of Fig.\,\ref{fig:fig4}\,top, with the 12/1\,ILR (with a remote contribution from the 8/1\,ILR).

In the SN, bin (c) in Fig.\,\ref{fig:fig3}, the two branches of Hercules scroll down, being associated now with levels II and III. The density of both decreases as the resonant domains of the corresponding 8/1 and 12/1\,ILRs diminish (panels II(c) and III(c) in Fig.\,\ref{fig:fig4}\,top). The remnants of these branches are seen in Fig.\,\ref{fig:fig3}(d), which are represented by lines I and II, respectively. For larger radii, the density of Hercules is very small, although we  can still observe the tenuous group around $V$=-70\,km\,s$^{-1}$ in panel (e) of Fig.\,\ref{fig:fig3}. The resonant pattern of Hercules is therefore consistent with the 8/1 and 12/1 ILRs, being the stronger and weaker branches, respectively.

\subsection{Sirius}

The Sirius stream is an outer structure with respect to the CR and we select four positive $V$-levels (V to VIII  in Fig.\,\ref{fig:fig3}) to characterize this group. The corresponding panels in Fig.\,\ref{fig:fig4}\,bottom show the presence of the OLRs, from -8/1 to -16/1, and higher. Contrasting with high-order ILRs, these resonances are dense and appear to be stuck to the CR, mainly in the V- and VI-levels in Fig.\,\ref{fig:fig4}\,bottom. This peculiar distribution of the OLRs may produce a dynamical phenomenon known as overlapping resonances, which commonly occurs in the vicinity of CRs. The overlap produces overdensity of objects which we observe in the $U$--$V$-distribution just above the CR in Fig.\,\ref{fig:fig3}. Indeed, Sirius already appears strongly  in the panel (b) between V- and VII-levels, has maximal density in (c) and (d) around the V-level and scrolls down following the CR in (e). The analysis of the corresponding panels in Fig.\,\ref{fig:fig4}\,bottom shows that Sirius, located very close to the large separatrix of the CR, is strongly influenced by these resonances, in such a way that, for large radii, the separation between them is no longer identifiable (e.g. bins (d)-(e) in  Fig.\,\ref{fig:fig3}). Moreover, the influence of the CR may explain the asymmetry with respect of the $U$-velocity contour of Sirius. Perturbations induced by the central bar could also break the symmetry of the $U$-distribution. Precise numerical simulations of the $U$--$V$ stellar distribution could shed light on this issue.



\subsection{Other structures}

A tenuous substructure below Hercules is observable in the planes (a)-(c) of Fig.\,\ref{fig:fig3}. According to panels I(b), II(a) and III(a) in Fig.\,\ref{fig:fig4}\,top, this group may be associated to the 6/1\,ILR. The weak output of the 6/1\,ILR is due to the fact that its order is not a multiple of 4, but it is still amazing to know that stars evolving inside this resonance visit the SN from the inner Galactic region (3--4\,kpc).

The strong 4/1\,OLR appears in panels VI(e), VII(d,e) and VIII(c,d) of Fig.\,\ref{fig:fig4}\,bottom. We see a small excess of stars in the corresponding upper parts of the lower planes of Fig.\,\ref{fig:fig3}. Next, we explain the relatively low population of this resonance due to the nature of their orbits.


\subsection{Resonant orbits}

\begin{figure}[h]
\begin{center}
\epsfig{figure=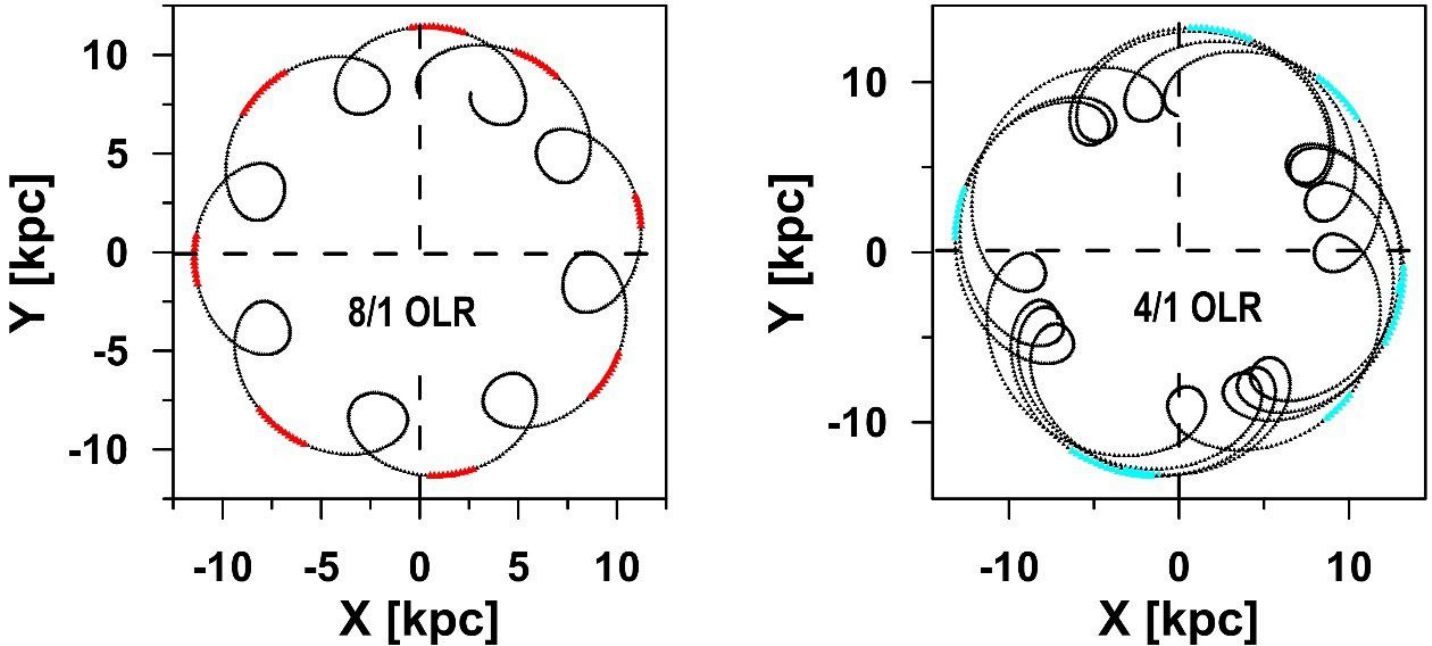,width=0.99\columnwidth ,angle=0}
\epsfig{figure=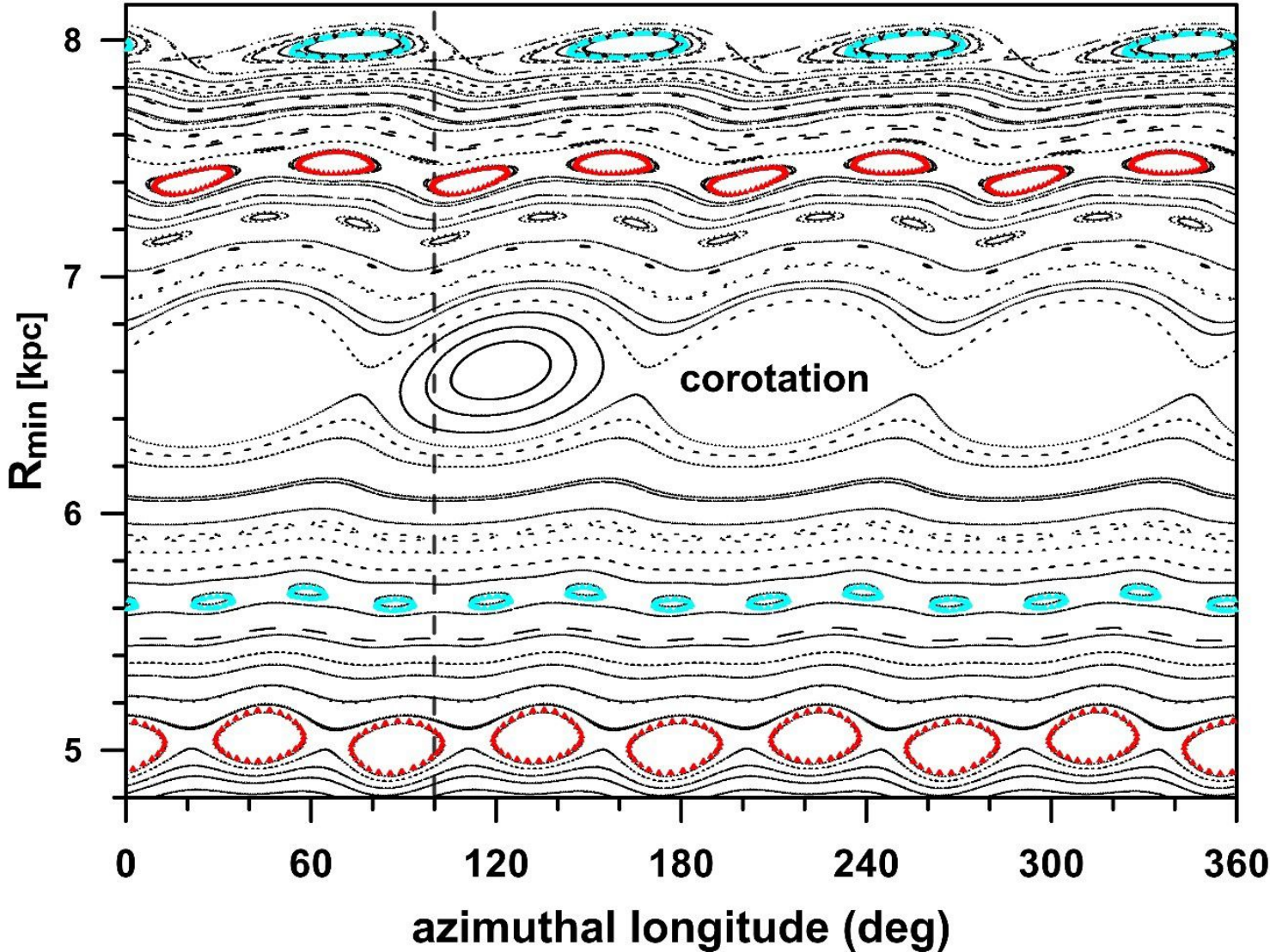,width=0.99\columnwidth ,angle=0}
\epsfig{figure=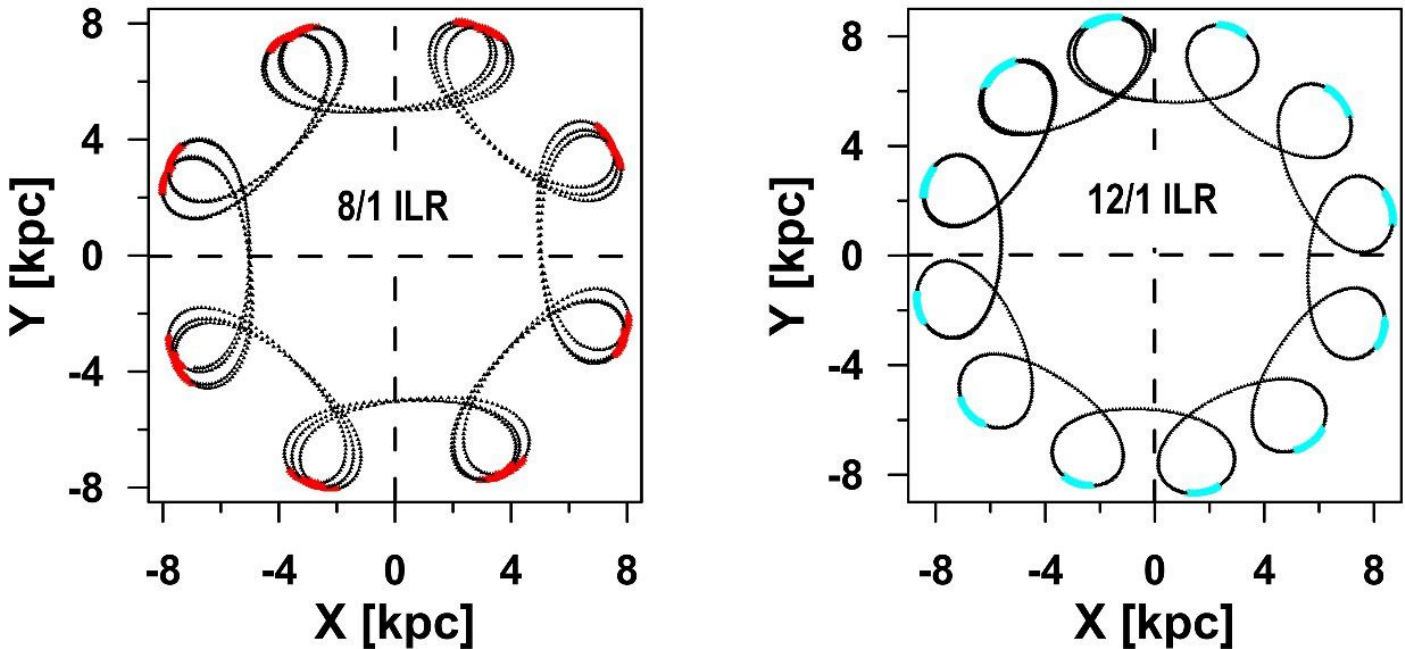,width=0.99\columnwidth ,angle=0}
\caption{Main panel: SoS ($p_R=0$ and $\dot p_R>0$), showing the corotation, ILRs and OLRs. The ILRs, 8/1 and 12/1, are red 8-island and cyan 12-island chains, respectively, below corotation; the OLRs, 8/1 and 4/1, are red 8-island and cyan 4-island chains above corotation. Top: two resonant orbits on the $X$--$Y$ planes: 8/1\,OLR (left corner) and 4/1\,OLR (right corner); colors identify the stretches of the orbits of lower velocities.  Bottom: same as on the top graphs, except for 8/1\,ILR (left corner) and 12/1\,ILR (right corner).
}
\label{fig:SOS}
\end{center}
\end{figure}

To complete the analysis, we construct the surfaces of section (SoS) along one energy level, which covers almost all possible regimes of motion in the SN. Figure \ref{fig:SOS} shows the SoS and four representative  orbits of stars that visit the SN. The SoS plot shows an 8-island chain (red) below the CR and the corresponding 8/1\,ILR orbit on the bottom-left panel. In this case, stars spend most of their time in the SN, enhancing the density in this region. We highlight the 12-island chain of the 12/1\,ILR by cyan color and show its orbit in bottom-right plot.

In the outer region, connected with Sirius, the 8/1\,OLR (red chain) appears above the CR in Fig.\,\ref{fig:SOS}, accompanied by the 12/1 and 16/1\,OLRs; its orbit is shown on the top-left panel. The 4-island cyan chain and its orbit on the top-right panel show that the 4/1\,OLR objects come from the Galaxy's confines, about 13--14\,kpc. The 4/1\,OLR is very strong, however, we do not observe the related high-density streams on the maps in Fig.\,\ref{fig:fig3}\,(d,e). In part, this is due to the fact that these objects spend most of their  time far away from the Sun, where their velocities are smaller. To obtain the metallicity of these stars, we select 531 objects from the Rave-TGAS surveys \citep{lindergrenEtal2016AA,kunderEtal2017AJ} in the interval 30$<$V$<$100\,km\,s$^{-1}$ and find that $\sim$40\% of this sample evolve inside the 4/1\,OLR. From these, the majority ($\sim$85\%) consists of metal-poor stars, with metallicity being $\sim$30\% in the range -1.35$<$[Fe/H]$<$-0.5 and $\sim$55\% in the range -0.5$<$[Fe/H]$<$0. This result reinforces that metal-poor stars come from the outer parts of the Galaxy \citep{QuillenEtal2018MNRAS,2018Hattori}.

\section{Summary}
\label{discussion}

We presented  a novel approach to explain the distribution of stars on the $U$--$V$ plane of the SN, based on a spiral arms dynamical model. Our basic assumptions are: i) the proximity between the spiral corotation radius and the solar circle, and ii) the dynamical stability of the Local Arm and the Sun. The suggestion that both the Local Arm and the Sun evolve inside the spiral CR yields natural constraints on the magnitude of the Galaxy's spiral pattern speed.  For the adopted rotation curve and solar position (Paper-I), we choose $\Omega_p = 28$\,km\,s$^{-1}$\,kpc$^{-1}$, which results in a corotation radius of $8.2$\,kpc.

The comparison of the stellar velocity distribution provided by Gaia DR2 with the dynamical maps of the $U$--$V$ plane shows that it is related to the CR and nearby Lindblad resonances. It is important to state that, in general,  resonances modify qualitatively the dynamics in their environment: they capture and trap stars inside the stable resonant zones, enhancing the density, and deplete regions close to saddle points and separatrices.   Corotation is special among the resonances, being stronger and having a wider zone of influence surrounded by many high-order resonances.

That is exactly what we observe on the $U$--$V$--plane (right) in Fig.\,\ref{fig:fig1}: the central region of enhanced density, known as main component, is the CR stable zone according to the dynamical map of the same plane (left). The notable moving groups, Coma-Berenices and Hyades-Pleiades, are inside the CR, together with the Sun, located at the origin of the plane. Immediately above the main component, for V$>$0, we can observe the Sirius group, whose origin is related to the overlapping high-order OLRs stuck to the CR.  
Finally, the strong 4/1\,OLR is responsible for distant stars to visit the SN.

In the region below the main component, at V$<$0, the nearby resonances are separated sufficiently on the $U$--$V$ plane to create several substructures well defined by Gaia DR2. The most prominent is the Hercules  group related mainly to the 8/1 and 12/1\,ILRs. The overdensities produced by the inner resonances are explained by the fact that the objects, coming from the inner Galaxy, spend longer times at the outer edges of their orbits in the SN, since their velocities are smaller there. This kinematics produces the effect of a bimodal distribution on the $U$--$V$ plane.

Despite the simplicity of our model, we are able to explain quantitatively the SN $U$--$V$ structure and its changes with different Galactic radii. The results presented here can  be considered as an additional test, which confirms the robustness of our model. They are also another strong indicator that the spiral corotation radius lies near the solar circle, since only in the neighborhood of the CR this abundance of Lindblad resonances can be observed. The bar's CR has a minor role, since, due to its spacial orientation,  the $L_4$--points of the bar are distant from the Sun, as shown in \cite{Michtchenkoetal2018AA}.

\begin{acknowledgments}
We acknowledge the anonymous referee for the detailed review and for the helpful suggestions which allowed us to improve the manuscript. This work was supported by the Brazilian CNPq and FAPESP (grants 2015/10577-9 and 2017/15893-1). This work has made use of the facilities of the Laboratory of Astroinformatics (IAG/USP, NAT/Unicsul), funded by FAPESP (grant 2009/54006-4) and INCT-A.
\end{acknowledgments}


\begin{thebibliography}{}
\expandafter\ifx\csname natexlab\endcsname\relax\def\natexlab#1{#1}\fi

\bibitem[{{Antoja} {et~al.}(2008){Antoja}, {Figueras}, {Fern{\'a}ndez}, \&
  {Torra}}]{antojaEtal2008AA}
{Antoja}, T., {Figueras}, F., {Fern{\'a}ndez}, D., \& {Torra}, J. 2008, \aap,
  490, 135

\bibitem[{{Antoja} {et~al.}(2011){Antoja}, {Figueras}, {Romero-G{\'o}mez},
  {Pichardo}, {Valenzuela}, \& {Moreno}}]{antojaEtal2011MNRAS}
{Antoja}, T., {Figueras}, F., {Romero-G{\'o}mez}, M., {et~al.} 2011, \mnras,
  418, 1423

\bibitem[{{Antoja} {et~al.}(2010){Antoja}, {Figueras}, {Torra}, {Valenzuela},
  \& {Pichardo}}]{antojaEtal2010LNEA}
{Antoja}, T., {Figueras}, F., {Torra}, J., {Valenzuela}, O., \& {Pichardo}, B.
  2010, Lecture Notes and Essays in Astrophysics, 4, 13

\bibitem[{{Antoja} {et~al.}(2009){Antoja}, {Valenzuela}, {Pichardo}, {Moreno},
  {Figueras}, \& {Fern{\'a}ndez}}]{antojaEtal2009ApJL}
{Antoja}, T., {Valenzuela}, O., {Pichardo}, B., {et~al.} 2009, \apjl, 700, L78

\bibitem[{{Antoja} {et~al.}(2014){Antoja}, {Helmi}, {Dehnen}, {Bienaym{\'e}},
  {Bland-Hawthorn}, {Famaey}, {Freeman}, {Gibson}, {Gilmore}, {Grebel},
  {Kordopatis}, {Kunder}, {Minchev}, {Munari}, {Navarro}, {Parker}, {Reid},
  {Seabroke}, {Siebert}, {Steinmetz}, {Watson}, {Wyse}, \&
  {Zwitter}}]{antojaEtal2014AA}
{Antoja}, T., {Helmi}, A., {Dehnen}, W., {et~al.} 2014, \aap, 563, A60

\bibitem[{{Antoja} {et~al.}(2018){Antoja}, {Helmi}, {Romero-Gomez}, {Katz},
  {Babusiaux}, {Drimmel}, {Evans}, {Figueras}, {Poggio}, {Reyle}, {Robin},
  {Seabroke}, \& {Soubiran}}]{antojaEtal2018arXiv}
{Antoja}, T., {Helmi}, A., {Romero-Gomez}, M., {et~al.} 2018, ArXiv e-prints,
  arXiv:1804.10196

\bibitem[{{Barros} {et~al.}(2016){Barros}, {L{\'e}pine}, \&
  {Dias}}]{BarrosLepine2016}
{Barros}, D.~A., {L{\'e}pine}, J.~R.~D., \& {Dias}, W.~S. 2016, \aap, 593, A108

\bibitem[{{Bensby} {et~al.}(2007){Bensby}, {Oey}, {Feltzing}, \&
  {Gustafsson}}]{bensbyEtal2007ApJL}
{Bensby}, T., {Oey}, M.~S., {Feltzing}, S., \& {Gustafsson}, B. 2007, \apjl,
  655, L89

\bibitem[{{Bovy}(2010)}]{bovy2010ApJ}
{Bovy}, J. 2010, \apj, 725, 1676

\bibitem[{{Bovy}(2015)}]{bovy2015ApJS}
---. 2015, \apjs, 216, 29

\bibitem[{{De Simone} {et~al.}(2004){De Simone}, {Wu}, \&
  {Tremaine}}]{desimoneEtal2004MNRAS}
{De Simone}, R., {Wu}, X., \& {Tremaine}, S. 2004, \mnras, 350, 627

\bibitem[{{Dehnen}(1999)}]{dehnen1999ApJL}
{Dehnen}, W. 1999, \apjl, 524, L35

\bibitem[{{Dehnen}(2000)}]{dehnen2000AJ}
---. 2000, \aj, 119, 800

\bibitem[{{Dias} \& {L{\'e}pine}(2005)}]{diasLepine2005ApJ}
{Dias}, W.~S., \& {L{\'e}pine}, J.~R.~D. 2005, ApJ, 629, 825

\bibitem[{{Famaey} {et~al.}(2008){Famaey}, {Siebert}, \&
  {Jorissen}}]{famaeyEtal2008AA}
{Famaey}, B., {Siebert}, A., \& {Jorissen}, A. 2008, \aap, 483, 453

\bibitem[{{Gaia Collaboration} {et~al.}(2018){Gaia Collaboration}, {Brown},
  {Vallenari}, {Prusti}, {de Bruijne}, {Babusiaux}, \&
  {Bailer-Jones}}]{Gaia2018A}
{Gaia Collaboration}, {Brown}, A.~G.~A., {Vallenari}, A., {et~al.} 2018, ArXiv
  e-prints, arXiv:1804.09365

\bibitem[{{Hattori} {et~al.}(2018){Hattori}, {Gouda}, {Yano}, {Sakai},
  {Tagawa}, {Baba}, \& {Kumamoto}}]{2018Hattori}
{Hattori}, K., {Gouda}, N., {Yano}, T., {et~al.} 2018, ArXiv e-prints,
  arXiv:1804.01920

\bibitem[{Hunt} {et~al.}(2018)]{Hunt2018arXiv180602832H}
{Hunt}, J.~A.~S., {Hong},~J., ,{Bovy},~J., {Kawata},~D. \& 	{Grand}, R.~J.~J. 2018,
{ArXiv e-prints}, arXiv:1806.02832

\bibitem[{{Junqueira} {et~al.}(2013){Junqueira}, {L{\'e}pine}, {Braga}, \&
  {Barros}}]{junqueiraEtal2013AA}
{Junqueira}, T.~C., {L{\'e}pine}, J.~R.~D., {Braga}, C.~A.~S., \& {Barros},
  D.~A. 2013, \aap, 550, A91

\bibitem[{{Katz} {et~al.}(2018){Katz}, {Sartoretti}, {Cropper}, {Panuzzo},
  {Seabroke}, {Viala}, {Benson}, {Blomme}, {Jasniewicz}, {Jean-Antoine},
  {Huckle}, {Smith}, {Baker}, {Crifo}, {Damerdji}, {David}, {Dolding},
  {Fr{\'e}mat}, {Gosset}, {Guerrier}, {Guy}, {Haigron}, {Jan{\ss}en},
  {Marchal}, {Plum}, {Soubiran}, {Th{\'e}venin}, {Ajaj}, {Allende Prieto},
  {Babusiaux}, {Boudreault}, {Chemin}, {Delle Luche}, {Fabre}, {Gueguen},
  {Hambly}, {Lasne}, {Meynadier}, {Pailler}, {Panem}, {Royer}, {Tauran},
  {Zurbach}, {Zwitter}, {Arenou}, {Bossini}, {Gomez}, {Lemaitre}, {Leclerc},
  {Morel}, {Munari}, {Turon}, {Vallenari}, \& {{\v Z}erjal}}]{KatzEtal2018A}
{Katz}, D., {Sartoretti}, P., {Cropper}, M., {et~al.} 2018, ArXiv e-prints,
  arXiv:1804.09372

\bibitem[{{Kunder} {et~al.}(2017){Kunder}, {Kordopatis}, {Steinmetz},
  {Zwitter}, {McMillan}, {Casagrande}, {Enke}, {Wojno}, {Valentini},
  {Chiappini}, {Matijevi{\v c}}, {Siviero}, {de Laverny}, {Recio-Blanco},
  {Bijaoui}, {Wyse}, {Binney}, {Grebel}, {Helmi}, {Jofre}, {Antoja}, {Gilmore},
  {Siebert}, {Famaey}, {Bienaym{\'e}}, {Gibson}, {Freeman}, {Navarro},
  {Munari}, {Seabroke}, {Anguiano}, {{\v Z}erjal}, {Minchev}, {Reid},
  {Bland-Hawthorn}, {Kos}, {Sharma}, {Watson}, {Parker}, {Scholz}, {Burton},
  {Cass}, {Hartley}, {Fiegert}, {Stupar}, {Ritter}, {Hawkins}, {Gerhard},
  {Chaplin}, {Davies}, {Elsworth}, {Lund}, {Miglio}, \&
  {Mosser}}]{kunderEtal2017AJ}
{Kunder}, A., {Kordopatis}, G., {Steinmetz}, M., {et~al.} 2017, \aj, 153, 75

\bibitem[{{L{\'e}pine} {et~al.}(2017){L{\'e}pine}, {Michtchenko}, {Barros}, \&
  {Vieira}}]{LepineEtal2017ApJ}
{L{\'e}pine}, J.~R.~D., {Michtchenko}, T.~A., {Barros}, D.~A., \& {Vieira},
  R.~S.~S. 2017, \apj, 843, 48

\bibitem[{{Lindegren} {et~al.}(2016){Lindegren}, {Lammers}, {Bastian},
  {Hern{\'a}ndez}, {Klioner}, {Hobbs}, {Bombrun}, {Michalik}, {Ramos-Lerate},
  {Butkevich}, {Comoretto}, {Joliet}, {Holl}, {Hutton}, {Parsons},
  {Steidelm{\"u}ller}, {Abbas}, {Altmann}, {Andrei}, {Anton}, {Bach},
  {Barache}, {Becciani}, {Berthier}, {Bianchi}, {Biermann}, {Bouquillon},
  {Bourda}, {Br{\"u}semeister}, {Bucciarelli}, {Busonero}, {Carlucci},
  {Casta{\~n}eda}, {Charlot}, {Clotet}, {Crosta}, {Davidson}, {de Felice},
  {Drimmel}, {Fabricius}, {Fienga}, {Figueras}, {Fraile}, {Gai}, {Garralda},
  {Geyer}, {Gonz{\'a}lez-Vidal}, {Guerra}, {Hambly}, {Hauser}, {Jordan},
  {Lattanzi}, {Lenhardt}, {Liao}, {L{\"o}ffler}, {McMillan}, {Mignard}, {Mora},
  {Morbidelli}, {Portell}, {Riva}, {Sarasso}, {Serraller}, {Siddiqui}, {Smart},
  {Spagna}, {Stampa}, {Steele}, {Taris}, {Torra}, {van Reeven}, {Vecchiato},
  {Zschocke}, {de Bruijne}, {Gracia}, {Raison}, {Lister}, {Marchant},
  {Messineo}, {Soffel}, {Osorio}, {de Torres}, \&
  {O'Mullane}}]{lindergrenEtal2016AA}
{Lindegren}, L., {Lammers}, U., {Bastian}, U., {et~al.} 2016, \aap, 595, A4

\bibitem[{{Lindegren} {et~al.}(2018){Lindegren}, {Hernandez}, {Bombrun},
  {Klioner}, {Bastian}, {Ramos-Lerate}, {de Torres}, {Steidelmuller},
  {Stephenson}, {Hobbs}, {Lammers}, {Biermann}, {Geyer}, {Hilger}, {Michalik},
  {Stampa}, {McMillan}, {Castaneda}, {Clotet}, {Comoretto}, {Davidson},
  {Fabricius}, {Gracia}, {Hambly}, {Hutton}, {Mora}, {Portell}, {van Leeuwen},
  {Abbas}, {Abreu}, {Altmann}, {Andrei}, {Anglada}, {Balaguer-Nunez},
  {Barache}, {Becciani}, {Bertone}, {Bianchi}, {Bouquillon}, {Bourda},
  {Brusemeister}, {Bucciarelli}, {Busonero}, {Buzzi}, {Cancelliere},
  {Carlucci}, {Charlot}, {Cheek}, {Crosta}, {Crowley}, {de Bruijne}, {de
  Felice}, {Drimmel}, {Esquej}, {Fienga}, {Fraile}, {Gai}, {Garralda},
  {Gonzalez-Vidal}, {Guerra}, {Hauser}, {Hofmann}, {Holl}, {Jordan},
  {Lattanzi}, {Lenhardt}, {Liao}, {Licata}, {Lister}, {Loffler}, {Marchant},
  {Martin-Fleitas}, {Messineo}, {Mignard}, {Morbidelli}, {Poggio}, {Riva},
  {Rowell}, {Salguero}, {Sarasso}, {Sciacca}, {Siddiqui}, {Smart}, {Spagna},
  {Steele}, {Taris}, {Torra}, {van Elteren}, {van Reeven}, \&
  {Vecchiato}}]{lindergrenEtal2018AA}
{Lindegren}, L., {Hernandez}, J., {Bombrun}, A., {et~al.} 2018, ArXiv e-prints,
  arXiv:1804.09366

\bibitem[{{Michtchenko} {et~al.}(2017)}]{Michtchenkoetal2017AA}{Michtchenko}, T.~A., {Vieira},  R.~S.~S., {Barros}, D.~A., \& {L{\'e}pine}, J.~R.~D. 2017, \aap, 597, id.A39

\bibitem[{{Michtchenko} {et~al.}(2018){Michtchenko}, {L{\'e}pine}, {Barros}, \&   {Vieira}}]{Michtchenkoetal2018AA}
{Michtchenko}, T.~A., {L{\'e}pine}, J.~R.~D., {Barros}, D.~A., \& {Vieira},
  R.~S.~S. 2018, \aap, 615, id.A10

\bibitem[{{Mishurov} \& {Zenina}(1999)}]{mishurovZenina1999AA}{Mishurov}, Y.~N., \& {Zenina}, I.~A. 1999, \aap, 341, 81

\bibitem[{{P{\'e}rez-Villegas} {et~al.}(2017){P{\'e}rez-Villegas}, {Portail},
  {Wegg}, \& {Gerhard}}]{perezvillegasEtal2017ApJL}
{P{\'e}rez-Villegas}, A., {Portail}, M., {Wegg}, C., \& {Gerhard}, O. 2017,
  \apjl, 840, L2

\bibitem[{{Quillen} \& {Minchev}(2005)}]{quillenMinchev2005AJ}
{Quillen}, A.~C., \& {Minchev}, I. 2005, \aj, 130, 576

\bibitem[{{Quillen} {et~al.}(2018{\natexlab{a}}){Quillen}, {Carrillo},
  {Anders}, {McMillan}, {Hilmi}, {Monari}, {Minchev}, {Chiappini}, {Khalatyan},
  \& {Steinmetz}}]{quillenEtal2018arXiv}
{Quillen}, A.~C., {Carrillo}, I., {Anders}, F., {et~al.} 2018{\natexlab{a}},
  ArXiv e-prints, arXiv:1805.10236

\bibitem[{{Quillen} {et~al.}(2018{\natexlab{b}}){Quillen}, {De Silva},
  {Sharma}, {Hayden}, {Freeman}, {Bland-Hawthorn}, {{\v Z}erjal}, {Asplund},
  {Buder}, {D'Orazi}, {Duong}, {Kos}, {Lin}, {Lind}, {Martell}, {Schlesinger},
  {Simpson}, {Zucker}, {Zwitter}, {Anguiano}, {Carollo}, {Casagrande}, {Cotar},
  {Cottrell}, {Ireland}, {Kafle}, {Horner}, {Lewis}, {Nataf}, {Ting}, {Watson},
  {Wittenmyer}, \& {Wyse}}]{QuillenEtal2018MNRAS}
{Quillen}, A.~C., {De Silva}, G., {Sharma}, S., {et~al.} 2018{\natexlab{b}},
  \mnras, arXiv:1802.02924

\bibitem[{{Sch{\"o}nrich} {et~al.}(2010){Sch{\"o}nrich}, {Binney}, \&
  {Dehnen}}]{schonrichBinneyDehnen2010MNRAS}
{Sch{\"o}nrich}, R., {Binney}, J., \& {Dehnen}, W. 2010, MNRAS, 403, 1829 (SBD)

\end{thebibliography}

\end{document}